\documentclass{article}
\usepackage{spconf,amsmath,graphicx,hyperref,caption}
\title{HD-PPT: Hierarchical Decoding of Content- and Prompt-Preference Tokens for Instruction-based TTS}

\name{
    Sihang Nie$^{1}$,
    Xiaofen Xing$^{1}$\sthanks{Corresponding author.}, 
    Jingyuan Xing$^{1}$, 
    Baiji Liu$^{1,2}$, 
    Xiangmin Xu$^{3,1*}$
}

\address{$^{1}$South China University of Technology, Guangzhou, China \\
  $^{2}$Guangzhou Quwan Network Technology, Guangzhou, China \\
  $^{3}$Foshan University, Foshan, China}

\usepackage{booktabs}
\usepackage{multirow}
\begin{document}
\maketitle

\begin{abstract}

Large Language Model (LLM)-based Text-to-Speech (TTS) models have already reached a high degree of naturalness. However, the precision control of TTS inference is still challenging. Although instruction-based Text-to-Speech (Instruct-TTS) models are proposed, these models still lack fine-grained control due to the modality gap between single-level text instructions and multilevel speech tokens. To address this limitation, we propose HD-PPT, a framework that transforms speech synthesis into a structured, hierarchical task. To enable fine-grained control, we introduce a novel speech codec to extract distinct prompt-preference and content-preference tokens from the complex speech tokens, supervised by automatic speech recognition (ASR) and cross-lingual audio-text pre-training (CLAP) objectives. To bridge the modality gap of these tokens, we propose a hierarchical decoding strategy, where the LLM generates tokens in a structured order: first semantic, then fine-grained style, and finally complete acoustic representation. Extensive experiments demonstrate that this hierarchical paradigm significantly improves instruction adherence and achieves state-of-the-art naturalness, validating our approach for precise and controllable speech synthesis. Audio samples are available at ~\url{https://xxh333.github.io/}.

\end{abstract}


\section{Introduction}
\label{sec:intro}

\begin{figure}[t!]
\begin{minipage}[hbtp]{1.0\linewidth}
  \centering
  \centerline{\includegraphics[width=8.6cm]{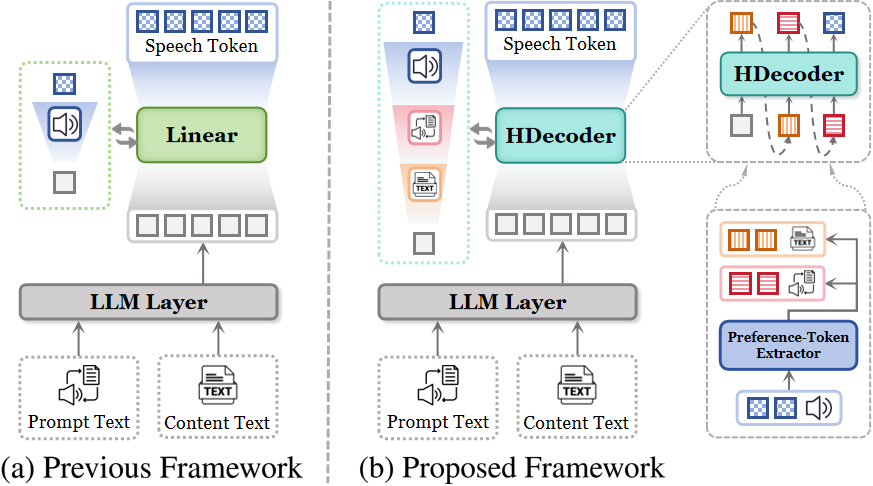}}
\end{minipage}
\caption{(a) shows a previous framework where LLM predicts a monolithic speech token sequence. (b) illustrates our proposed HD-PPT, which guides the LLM to model speech hierarchically by extracting content- and prompt-preference tokens from the speech tokens and jointly modeling them.}

\label{fig:mot}
\end{figure}

Recently, the naturalness of TTS models has achieved substantial progress~\cite{cosy1, sparktts, minimax}. However, the precise control of human-like speech synthesis remains a central challenge in Text-to-Speech (TTS), with expressive attributes like prosody, emotion, and timbre. To solve this problem, the instruction-based Text-to-Speech (Instruct-TTS) paradigm is proposed, aiming to generate high-quality speech that precisely adheres to descriptive natural language prompts~\cite{prompttts, instructtts, emovoice}.

Current approaches fall largely into two main categories: explicit style encoding methods~\cite{prompttts, instructtts, prompttts2, promptstyle, controlspeech} and Large Language Model (LLM)-driven methods~\cite{cosy1, emovoice, cosy2, textrolspeech, voxinstruct}. Although explicit style encoding methods can achieve basic prompt control, they are limited by their inefficient structure and coarse-grained control. In contrast, LLM-based methods offer a more flexible architecture and stronger control for interpreting nuanced textual instructions. Despite their promise, these methods often struggle with precision and robustness, particularly when faced with complex or subtle prompts. This is primarily because they directly map the style information from the text instructions onto the speech tokens, making fine-grained control difficult, as shown in Fig.~\ref{fig:mot}(a). In addition, they treat speech tokens as a monolithic sequence. We posit that this limitation stems from a fundamental hierarchical mismatch: they attempt to map a single-level text instruction directly onto multilevel speech tokens. This approach overlooks the inherently hierarchical nature of speech, which involves three types of information: linguistic, paralinguistic, and extralinguistic, corresponding to spoken content, prosody/emotion, and speaker/scenario, respectively~\cite{s1}.

\begin{figure*}[htbp]
\hspace*{-0.6cm}
\begin{minipage}[b]{0.48\linewidth}
  \centerline{\includegraphics[width=7.4cm]{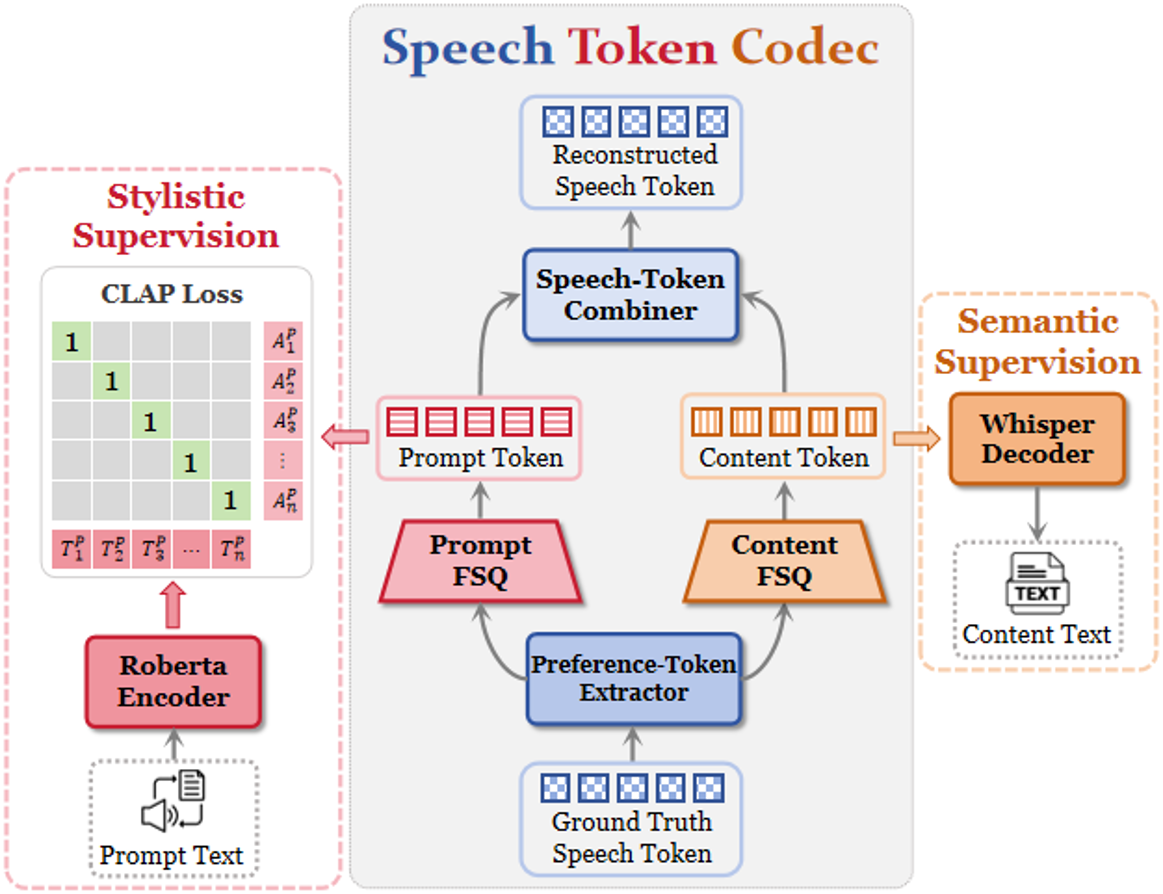}}
  \centerline{(a) Speech Token Codec}\medskip
\end{minipage}
\hspace*{0.2cm}
\begin{minipage}[b]{0.48\linewidth}
  \centerline{\includegraphics[width=10cm]{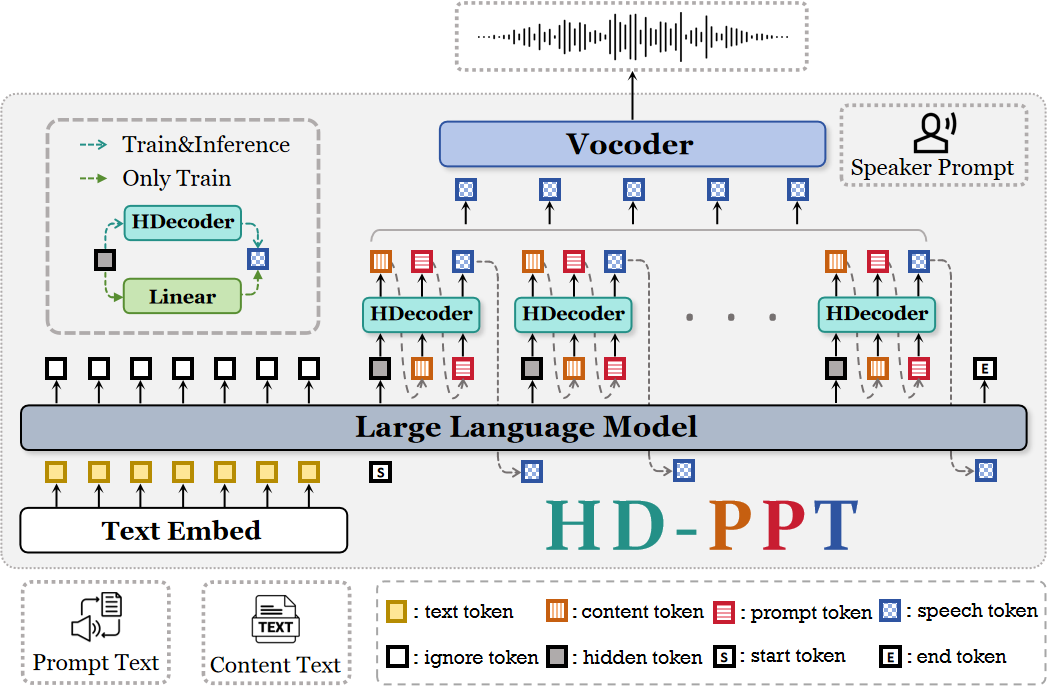}}
  \centerline{(b) An overview of the proposed HD-PPT}\medskip
\end{minipage}
\caption{Figure (a) illustrates the speech token codec, which extracts content- and prompt-preference tokens from speech tokens. Figure (b) shows the overall architecture of HD-PPT, comprising the hierarchical LLM and a subsequent vocoder.}
\label{fig:mod}
\end{figure*}

To resolve this hierarchical mismatch, we reframe the synthesis task from monolithic generation to a structured process. We propose \textbf{HD-PPT},  a framework for \textbf{H}ierarchical \textbf{D}ecoding of Content- and \textbf{P}rompt-\textbf{P}reference \textbf{T}okens for Instruct-TTS. As illustrated in Fig.~\ref{fig:mot}(b), our approach is founded on two key innovations designed to bridge the gap between instruction and audio. To enable fine-grained control, we introduce a novel speech token codec. Jointly supervised by automatic speech recognition (ASR) and cross-lingual audio-text pre-training (CLAP)~\cite{clap}, it distinguishes between prompt-preference tokens to capture fine-grained style and content-preference tokens to anchor semantics. To bridge the hierarchical gap, we design a hierarchical decoding strategy. This guides the LLM to generate these representations sequentially: first establishing the semantic foundation, then layering stylistic details, and finally rendering the complete acoustic representation. This structured generation process dramatically enhances the model's ability to execute instructions with precision and fidelity.

In summary, our main contributions are as follows.
1) A novel speech codec to extract and differentiate content- and prompt-preference tokens, providing a fine-grained intermediate modeling target for the LLM.
2) A hierarchical decoding strategy that aligns generation with the intrinsic structure of speech, guiding the LLM to render audio hierarchically to improve complex instruction execution.
3) Extensive validation of our method's effectiveness, demonstrating state-of-the-art performance in both naturalness and control accuracy.

\section{METHODOLOGY}
\label{sec:method}

The HD-PPT framework consists of three main components, as depicted in Fig.~\ref{fig:mot}: 1) a speech token codec designed to extract content- and prompt-preference tokens from the speech token; 2) an LLM with a hierarchical decoder, which receives natural language instructions and generates various token sequences in a structured manner; and 3) a vocoder, which synthesizes the final waveform from the generated speech tokens and speaker embeddings. The core design principle is to transform speech synthesis from predicting an undifferentiated acoustic sequence into a structured hierarchical generation process.

\subsection{Speech Token Codec with Content- and Prompt-Preference Token Extraction}
\label{ssec:codec}
To effectively extract fine-grained preference representations from speech, we designed a speech token codec based on finite-scalar quantization (FSQ)~\cite{fsq}, as illustrated in Fig.~\ref{fig:mod}(a). The model is optimized via a combination of a reconstruction loss and two auxiliary supervision tasks.

The codec employs a transformer-based architecture. A preference token extractor first encodes the input speech tokens (from the pre-trained CosyVoice2~\cite{cosy2} tokenizer) into a continuous representation $Z$. Subsequently, two independent FSQ modules quantize $Z$ into distinct, discrete preference tokens. Finally, a causal transformer-based speech token combiner fuses these preference tokens to reconstruct the original speech tokens. This causal design enforces temporal alignment between the representations. In addition, slight random noise is injected during training to enhance robustness.

To ensure that the preference tokens capture distinct speech attributes, we impose specific supervision mechanisms. The content-preference tokens are supervised by an ASR task, using a Whisper-Small decoder~\cite{whisper} to predict text, thus exposing them to semantic information. The prompt-preference tokens are supervised by a CLAP-based contrastive loss~\cite{clap} to capture prosody and emotion. A cross-attention module maps these tokens to a fixed-length embedding. This embedding is trained to maximize cosine similarity with the text embedding of the corresponding prompt (from a pre-trained RoBERTa-base model~\cite{roberta}), while minimizing similarity to embeddings of noncorresponding prompts. This objective compels the prompt-preference tokens to encode fine-grained stylistic attributes correlated with the prompt.

The total loss is a weighted sum of the reconstruction, ASR, and CLAP losses:
\begin{equation}
L_{total}=L_{rec}+\lambda_{asr}L_{asr}+\lambda_{clap}L_{clap}
\end{equation}

where $L_{rec}$ is the cross-entropy loss for reconstruction, and the weights $\lambda_{asr}$ and $\lambda_{clap}$ are set to 2.0 and 0.8, respectively. Through this joint optimization strategy, the codec effectively learns to extract different preference representations tailored for hierarchical synthesis.

\subsection{LLM's Hierarchical Decoding}
\label{ssec:llm}
With the preference speech tokens established, we leverage an LLM to generate them from textual instructions. We chose Qwen2.5-0.5B~\cite{qwen} as backbone, paired with a lightweight transformer decoder to perform hierarchical generation.

The LLM auto-regressively generates a sequence of hidden states $T_h$ based on input text $T_t$. At each step, the hidden state is fed into the lightweight hierarchical decoder to sequentially predict the tokens. Assume that the content-preference tokens, prompt-preference tokens, and speech tokens are $T_c$, $T_p$, and $T_s$, respectively. Additionally, let $\theta_{LM}$, $\theta_{HD}$ represent the parameters of the LLM and the decoder. The generation process at step $j$ is as follows:

\textbf{Content Foundation}. The modified LLM generates a hidden state $T_{h,j}$, which is then passed to the hierarchical decoder to produce the content-preference token $T_{c,j}$. Note that $T_{s,:j}$ denotes the sequence history of speech tokens prior to step $j$. This step establishes a semantic basis. 
\begin{gather}
p(T_{h,j}|T_t;\theta_{LM})=p(T_{h,j}|T_t,T_{s,:j}) \\
p(T_{c,j}|T_t;\theta_{LM},\theta_{HD})=p(T_{c,j}|T_{h,j})
\end{gather}

\textbf{Style Rendering}. Subsequently, the model renders stylistic attributes. The prompt-preference token $T_{p,j}$ is predicted by the decoder, conditioned on both the hidden state $T_{h,j}$ and the newly generated content token $T_{c,j}$:
\begin{equation}
p(T_{p,j}|T_t;\theta_{LM},\theta_{HD})=p(T_{p,j}|T_{h,j},T_{c,j})
\end{equation}

\textbf{Final Token Generation}. Finally, with both semantic and stylistic foundations in place, the complete speech token $T_{s,j}$ is predicted by fusing all prior information:
\begin{equation}
p(T_{s,j}|T_t;\theta_{LM},\theta_{HD})=p(T_{s,j}|T_{h,j},T_{c,j},T_{p,j})
\end{equation}

The resulting speech token $T_{s,j}$ is then fed back to the modified LLM to generate the next hidden state $T_{h,j+1}$ for the next timestep.

To ensure that the model robustly learns this hierarchical process, we employ two regularization strategies during training. First, we introduce stochasticity by probabilistically masking the hidden states and prompt tokens, and by concatenating token logits with the token embeddings as input to the lightweight decoder. These interventions compel the model to integrate the information from all available sources rather than relying on a single one. Second, as shown in Fig.~\ref{fig:mod}(b), an auxiliary linear layer is added to directly project the LLM's hidden states into the speech tokens, ensuring that its internal representations remain acoustically grounded.

\section{EXPERIMENTS}
\label{sec:experiments}

\subsection{Experimental Setup}
\label{ssec:setup}
\hspace{5.3mm}\textit{1) Datasets and Baselines.} We conducted experiments on two public datasets to ensure a comprehensive evaluation: TextrolSpeech~\cite{textrolspeech} for fine-grained style control and EmoVoice-DB~\cite{emovoice} for emotional control. All audio was resampled to 24kHz. We compared HD-PPT against two categories of baselines: 1) \textbf{Explicit style encoding}: PromptTTS~\cite{prompttts} and PromptStyle~\cite{promptstyle}; and 2) \textbf{LLM-driven}: CosyVoice~\cite{cosy1}, EmoVoice-PP~\cite{emovoice}, and our main baseline, CosyVoice2~\cite{cosy2}.

\textit{2) Evaluation metrics}. Our evaluation employed a combination of subjective and objective metrics. For subjective tests, 18 participants rated speech naturalness (MOS-N) and stylistic consistency (MOS-S) on a 5-point Likert scale. The evaluation set comprised 18 samples curated to cover diverse emotion categories and instructions, ensuring a balanced assessment. Objectively, We used the CV3-Eval toolkit~\cite{cosy3, dro} to obtain perceptual quality through the Deep Noise Suppression Mean Opinion Score (DNSMOS) and the word error rate (WER). Additionally, emotional similarity (EMO-SIM) was calculated via cosine similarity of emotion2vec-plus-large~\cite{emotion2vec} features between real and synthesized audio.
    
\textit{3) Implementation details}. We first trained our speech token codec, which consists of a 5-layer conformer~\cite{conformer} extractor and a 4-layer causal transformer combiner. The FSQ codebook sizes for the prompt- and content-preference tokens were set to 64 and 1296, respectively, both operating at a rate of 25Hz. The codec was trained for 50 epochs on 4 NVIDIA 4090 GPUs using the AdamW optimizer with a learning rate of $1\times10^{-4}$. Following this, we trained the modified LLM, which uses Qwen2.5-0.5B as its backbone. For this model, we employed a lightweight 2-layer auto-regressive transformer with a fixed length of 3 as the hierarchical Decoder. The LLM was trained for 16 epochs on the same hardware using the AdamW optimizer, but with a learning rate of $1\times10^{-5}$. For the final audio generation, we used the official pre-trained vocoder from CosyVoice2~\cite{cosy2}, which combines a flow-matching model~\cite{flow} and HifiGAN~\cite{hifigan}.

\begin{table}[t]
    \centering
    \captionsetup{margin={2em, -2em}} 
    \caption{Subjective and objective comparison on test sets.}
    \label{tab:main_results}
    \makebox[\linewidth][l]{
        \resizebox{1.17\linewidth}{!}{
            \begin{tabular}{@{}lcccccc@{}}
            \toprule
            \multirow{2}{*}{\textbf{Model}} & \multicolumn{2}{c}{\textbf{Subjective}} & \multicolumn{3}{c}{\textbf{Objective}} \\
            \cmidrule(lr){2-3} \cmidrule(lr){4-6}
            & \textbf{MOS-N~↑} & \textbf{MOS-S~↑} & \textbf{DNSMOS~↑} & \textbf{EMO-SIM~↑} & \textbf{WER~↓} \\ 
            \midrule
            PromptStyle & 2.674 $\pm$ 0.145 & 2.420 $\pm$ 0.147 & 3.68 & 0.529 & 17.92\% \\
            PromptTTS & 2.920 $\pm$ 0.137 & 2.601 $\pm$ 0.148 & 3.65 & 0.588 & \textbf{4.38\%} \\
            CosyVoice & 3.240 $\pm$ 0.138 & 3.028 $\pm$ 0.149 & 3.77 & 0.635 & 6.10\% \\
            CosyVoice2 & 3.920 $\pm$ 0.112 & 3.885 $\pm$ 0.116 & 3.83 & 0.714 & 5.71\% \\
            EmoVoice-PP & 3.694 $\pm$ 0.123 & 3.594 $\pm$ 0.128 & \textbf{3.87} & 0.613 & 8.56\% \\
            \midrule
            \textbf{HD-PPT (Ours)} & \textbf{4.108 $\pm$ 0.105} & \textbf{4.167 $\pm$ 0.103} & 3.84 & \textbf{0.753} & 5.18\% \\ 
            \bottomrule
            \end{tabular}
        }
    }
\end{table}

\subsection{Experimental Results}
\label{ssec:results}
\subsubsection{Comparison with Baselines}
\label{sssec:comparison}

Table~\ref{tab:main_results} presents a comprehensive comparison between HD-PPT and the five baseline models on the combined test sets of TextrolSpeech and EmoVoice-DB. HD-PPT achieves superior performance across the board. In subjective tests, it received the highest MOS-N and MOS-S scores, which prove its excellent naturalness and stylistic consistency. Objectively, it achieved the best EMO-SIM score for controllable emotional expression. These high scores directly validate that our hierarchical structure improved instruction adherence and stylistic control. Furthermore, HD-PPT also achieved a competitive DNSMOS and the second lowest WER, demonstrating its ability to generate high-fidelity and intelligible speech.

\subsubsection{Ablation on Preference Tokens}
\label{sssec:ablation_codec}
To validate the efficacy of preference tokens, we conducted ablation experiments across four variants: 1) \textbf{w/o Content-Pref.}: removing content-preference tokens from the decoding process; 2) \textbf{w/o Prompt-Pref.}: removing prompt-preference tokens; 3) \textbf{w/o Dual-Pref.}: bypassing both preference tokens; and 4) \textbf{w/o Instruct Text}: generating speech without the style prompt. As shown in Table~\ref{tab:ablation_tokens}, removing either preference token led to a performance drop. The removal of content-preference tokens caused a significant increase in WER, highlighting their role in maintaining semantic integrity. The absence of prompt-preference tokens led to a notable decrease in EMO-SIM, underscoring their necessity for stylistic nuances. When both were removed, all metrics degraded, confirming the importance of our structured intermediate representations. Furthermore, the drastic drop in EMO-SIM without instruction text proves that the model's stylistic control is directly derived from the prompt rather than dataset bias.

\begin{table}[t]
    \centering
    \captionsetup{margin={1em, -2em}}
    \caption{Ablation on preference tokens.}
    \label{tab:ablation_tokens}
    \makebox[\linewidth][r]{
        \resizebox{0.84\linewidth}{!}{
            \begin{tabular}{@{}lccc@{}}
            \toprule
            \textbf{Model} & \textbf{DNSMOS ↑} & \textbf{EMO-SIM ↑} & \textbf{WER ↓} \\ \midrule
            w/o Content-Pref. & 3.76 & 0.742 & 8.04\% \\
            w/o Prompt-Pref. & 3.76 & 0.728 & 5.49\% \\
            w/o Dual-Pref. & 3.73 & 0.716 & 10.10\% \\
            w/o Instruct Text & 3.78 & 0.605 & 5.44\% \\ \midrule
            \textbf{Proposed} & \textbf{3.84} & \textbf{0.753} & \textbf{5.18\%} \\ \bottomrule
            \end{tabular}
        } 
    }
    
    \vspace{1em}
    \captionsetup{margin={1em, -2em}} 
    \caption{Ablation on hierarchical decoding strategy.}
    \label{tab:ablation_decoding}
    \makebox[\linewidth][r]{
        \resizebox{0.84\linewidth}{!}{
            \begin{tabular}{@{}lccc@{}}
            \toprule
            \textbf{Model} & \textbf{DNSMOS ↑} & \textbf{EMO-SIM ↑} & \textbf{WER ↓} \\ \midrule
            Parallel & 3.76 & 0.736 & 5.99\% \\
            Single-step & 3.80 & 0.713 & 5.93\% \\ \midrule
            \textbf{Hierarchical} & \textbf{3.84} & \textbf{0.753} & \textbf{5.18\%} \\ \bottomrule
            \end{tabular}
        }
    }
\end{table}

\subsubsection{Ablation on Hierarchical Decoding Strategy}
\label{sssec:ablation_decode}
We evaluated our hierarchical decoding strategy against two alternatives: 1) \textbf{Parallel}: predicting all three token types (content, prompt, speech) simultaneously from the LLM's hidden state. 2) \textbf{Single-step}: directly predicting the final speech tokens, bypassing the intermediate preference tokens. Results in Table~\ref{tab:ablation_decoding} show that our hierarchical approach outperforms both. The suboptimal results of the parallel approach demonstrated that an explicit conditional dependency is needed for effective output structuring. The weaker performance of the single-step model further affirmed the need for structured intermediate representations. This validates that sequential, layer-by-layer decoding is essential for precise control. Regarding latency on an NVIDIA 4090, the Real-Time Factor (RTF) increased from 0.711 (single-step) to 0.952 (ours). We consider this moderate cost acceptable given the significant gains in control precision.

\section{CONCLUSION}
\label{sec:conclusion}

In this paper, we introduce HD-PPT, a novel framework for Instruct-TTS that resolves the hierarchical mismatch between textual instructions and speech signals. By employing a specialized codec to extract dual preference tokens from speech tokens and a hierarchical decoding strategy to generate them sequentially, our method significantly enhances fine-grained control and expressiveness. Extensive experiments demonstrated that HD-PPT outperforms state-of-the-art baselines in both instruction adherence and speech naturalness. We acknowledge that the multi-component complexity poses challenges for low-resource language adaptation. Consequently, future work will integrate Reinforcement Learning to mitigate this difficulty. Furthermore, we aim to extend the framework to advanced emotional speech synthesis, enabling more fine-grained affective generation.

\section{ACKNOWLEDGEMENT}
\sloppy
This work was supported by the Guangdong Basic and Applied Basic Research Foundation under Grant 2025A1515011203, the Guangdong Provincial Key Laboratory of Human Digital Twin under Grant 2022B1212010004, the Key R\&D and Achievement Transformation Program of Inner Mongolia Autonomous Region (2025YFHH0014).

\begin{small}
\bibliographystyle{IEEEbib}
\bibliography{refs,strings}
\end{small}

\end{document}